\documentclass[%
 reprint,
amsmath,amssymb,
aps,
pub,
]{revtex4-2}

\usepackage{graphicx}% Include figure files
\usepackage{dcolumn}% Align table columns on the decimal point
\usepackage{bm}% bold math
\usepackage{color}
\usepackage{amsmath}

\begin{document}

\preprint{APS/123-QED}

\title{Exploring the role of four-phonon scattering in the lattice thermal transport of LaMoN$_3$} % Force line breaks with \\

\author{Manjari Jain$^*$}
\author{Sanchi Monga} 
\author{Saswata Bhattacharya$^*$}%
\affiliation{%
 Department of Physics, Indian Institute of Technology Delhi, New Delhi 110016, India\\
}%
\date{\today}% It is always \today, today,

\begin{abstract}
In this work, we systematically investigate the lattice thermal conductivity ($\kappa_L$) of LaMoN$_3$ in the $C$2/$c$ and $R$3$c$ phases using first-principles calculations combined with the Boltzmann transport equation. In the $C$2/$c$ phase, $\kappa_L$ exhibits strong anisotropy, with values of 0.75 W/mK, 1.89 W/mK, and 0.82 W/mK along the a, b, and c axes, respectively, at 300 K. In contrast, the $R$3$c$ phase shows nearly isotropic thermal conductivity, with values of 6.28 W/mK, 7.05 W/mK, and 7.31 W/mK along the a, b, and c directions. In both phases, acoustic phonons dominate thermal transport. However, in the $C$2/$c$ phase, the absence of an acoustic-optical gap results in increased three-phonon scattering leading to smaller values of $\kappa_L$. Additionally, four-phonon scattering plays a dominant role in the C2/c phase, reducing $\kappa_L$ by approximately 96\%, whereas in the $R3c$ phase, it leads to a smaller but still significant reduction of ~50\%. These results highlight the critical role of four-phonon interactions in determining the thermal transport properties of LaMoN$_3$ and reveal the stark contrast in thermal conductivity between its two structural phases.
\end{abstract}
\maketitle
 
%\tableofcontents

\section{\label{sec:level1}Introduction}

Perovskites of the ABX$_3$ type have emerged as a prominent class of materials, distinguished by a three-dimensional network of corner-sharing BX$_6$ octahedra, with the A-site cation nestled in the cavities formed by these octahedra~\cite{Shi2020, D0TC01484B}. These materials exhibit many remarkable properties, including ferroelectricity, ferromagnetism, piezoelectricity, and optoelectronic features, rendering them valuable for scientific research and technological applications~\cite{Lee2015, Tang2015, Kojima2009}. For instance, due to their significant piezoelectric responses, oxide perovskites such as (Pb, Ba)TiO$_3$ have been extensively utilized in electrochemical cells, ceramic capacitors, and microelectromechanical actuators~\cite{Dimos1998, Muralt2009, Papac2021, Boris2011}. More recently, halide perovskites (APbX$_3$) have garnered considerable attention for their exceptional optoelectronic properties, propelling perovskite solar cells to achieve power conversion efficiencies exceeding 26\%~\cite{Kim2020, Tiwari2021, Gill2024, Wang2017}.

Despite the extensive research on oxide and halide perovskites, nitride compounds have gained increasing attention due to the moderate electronegativity of nitrogen ($\chi_N$  = 3.0) and their mixed covalent/ionic bonding characteristics~\cite{Zakutayev2016, Jain2023Oxynitride}. Nitrides exhibit superior solar absorption and electrical transport capabilities compared to their oxide counterparts, yet they remain largely unexplored. Some theoretical investigations have attempted to substitute the X-site anion from Group VI/VII elements with Group V elements (i.e., X = N) to explore the feasibility of nitride perovskites~\cite{SarmientoPerez2015}. For example, ABN$_3$ (A = La, Ce, Eu, Yb; B = W, Re) materials were predicted to be thermodynamically stable, with lanthanum tungsten nitride (LaWN$_3$) being synthesized~\cite{Talley2021}. CeWN$_3$ and CeMoN$_3$ were discovered through high-throughput computational screening and thin film growth techniques~\cite{Sherbondy2022}.

Among these nitride perovskites, LaWN$_3$ was anticipated to possess a modest band gap [hybrid functional (HSE06) result: 1.72 eV] and exhibit ferroelectricity with a spontaneous polarization of about 66 $\mu$C/cm$^2$~\cite{PhysRevB.95.014111}. However, due to the spatially extended 5d orbitals, LaWN$_3$ has a small band gap, preventing experimental confirmation of its ferroelectricity. Most ferroelectric perovskites are based on 3d metal oxides, such as Pb(Zr, Ti)O$_3$ and BaTiO$_3$ with more localized d orbitals resulting in larger band gaps. Consequently, there is a need to find more ferroelectric nitride perovskites, particularly those with wider band gaps. It is reasonable to consider using 3d or 4d cations instead of the B-site W$^{6+}$ cation, such as in LaMoN$_3$ with a HSE06 band gap of 1.98 eV. The ground state structure of LaMoN$_3$ was predicted to be a nonpolar, nonperovskite $C$2/$c$  phase~\cite{doi:10.1021/acs.chemmater.5b02026}. Subsequent studies reported a structural change in LaMoN$_3$ from the nonpolar, nonperovskite $C$2/$c$ phase to the ferroelectric perovskite $R$3$c$  phase induced by pressure~\cite{PhysRevB.102.180103}. Furthermore, LaMoN$_3$ may exhibit better ferroelectric properties than LaWN$_3$. Nonetheless, unlike oxide and halide perovskites, the physical properties of nitride perovskites, particularly those related to heat transport, remain largely unexplored.

The thermal conductivity of a material, a measure of its ability to transport heat under a limited temperature gradient, is crucial for many modern technologies, including photovoltaics, transistors, and thermoelectric devices~\cite{Bell2008}. Extensive research on the lattice thermal conductivity ($\kappa$$_L$) in perovskite materials demonstrates the relevance of systematic studies in this area~\cite{Wu2018, Zhao2021}. For example, halide perovskites have ultralow lattice thermal conductivity [$<$0.65 W/mK], making them potential candidates for significant thermoelectric applications~\cite{Haque2020, Liu2019}. In contrast, oxide perovskites typically exhibit better thermal conductivity [$\sim$5–10 W/mK] and good mechanical stability at high temperatures (1000 K)~\cite{Roekeghem2016}.

Density functional theory (DFT) based thermal conductivity calculations have gained popularity due to their low computational cost and generally good agreement with experimental $\kappa$$_L$  for various systems~\cite{10.1063/1.2822891, PhysRevLett.106.045901}. These calculations primarily consider the lowest-order intrinsic phonon scattering events involving three-phonons to reduce computational cost. However, recent methodological developments have incorporated higher-order anharmonic terms involving four-phonon scattering events, significantly reducing the error in $\kappa$$_L$  estimations~\cite{PhysRevB.93.045202, PhysRevB.96.161201, PhysRevB.97.045202}.

In this study, we employ DFT combined with the Boltzmann transport equation (BTE) to systematically investigate the heat transport properties of the nitride LaMoN$_3$ in its $C$2/$c$ and $R$3$c$ phases. To elucidate the phonon-related mechanisms in LaMoN$_3$, a comprehensive analysis of phonon dispersion, Grüneisen parameter, group velocity, phonon lifetime, and lattice thermal conductivity is conducted. This study aims to address the existing gap in the literature concerning the thermal properties of nitride perovskites and to provide valuable insights into their potential applications in advanced technological fields.

\section{\label{sec:level2}Methodology}
DFT calculations were performed using the Vienna $ab$ $initio$ simulation package (VASP) with projector-augmented wave (PAW) pseudopotentials~\cite{Kresse1996, Kresse1996b, Blochl1994}. The electronic exchange and correlation interactions were addressed using the generalized gradient approximation (GGA) with the Perdew-Burke-Ernzerhof (PBE) functional~\cite{perdew1996generalized}. A plane wave cutoff energy of 600 eV was chosen, and structural relaxations were conducted with a 6$\times$6$\times$6 $\Gamma$-centered $ k$-point mesh, achieving convergence of Hellmann-Feynman forces to within 10$^{-3}$ eV/\AA. To ensure accuracy, an energy convergence threshold of 10$^{-8}$ eV was applied for phonon computations.

Phonon dispersion relations were calculated using the PHONOPY code~\cite{Togo2015} with a 2$\times$2$\times$2 supercell through the finite displacement method~\cite{Esfarjani2008}. This approach allowed for the computation of phonon band structures, phonon density of states, and mode-resolved group velocities at arbitrary $q$ vectors~\cite{Wang2010}. The lattice thermal conductivity $\kappa_L$ at temperature T was determined by solving the BTE for phonons using the SHENGBTE code~\cite{Li2014, Han2022}. The thermal conductivity is expressed as:

\begin{equation}
\kappa_L^{\alpha \beta} = \frac{\hbar^2}{k_B T^2 N \Omega} \sum_{\lambda} f_0 (f_0 + 1) (\omega_{\lambda})^2 v_{\lambda}^{\alpha} v_{\lambda}^{\beta} \tau_{\lambda},
\end{equation}

where $\Omega$ is the volume of the unit cell, N denotes the number of $ k$-points, and f$_0$ is the Bose-Einstein distribution function. Here, \(\omega_{\lambda}\) is the angular frequency of the phonon mode \(\lambda\), \(v_{\lambda}^{\alpha}\) and \(v_{\lambda}^{\beta}\) represent the components of the group velocity, and \(\tau_{\lambda}\) denotes the phonon lifetime.

The interatomic force constants (IFCs) up to the fourth order were computed using the finite-difference supercell method. For third-order IFCs, a 2$\times$2$\times$2 supercell was employed, considering interactions up to the sixth nearest-neighbor (NN) atoms with an extension module in SHENGBTE. Fourth-order IFCs were calculated using the unit cell with interactions up to 4 NN atoms. All supercell-based calculations utilized a well-converged $k$-point mesh and adhered to a rigorous energy convergence threshold of 10$^{-8}$ eV. The three and four-phonon scattering rates were also determined with SHENGBTE, using 12$\times$12$\times$12 and 4$\times$4$\times$4 $q$-point grids, respectively, and a scalebroad of 0.5. 

\begin{table}[b]
\caption{\label{tab:table1}%
Lattice parameters of LaMoN$_3$ in the $C$2/$c$ and $R$3$c$ phase.}
\begin{ruledtabular}
% Set the row separation
\renewcommand{\arraystretch}{1.5} % Default value is 1
\begin{tabular}{lccc}
\textrm{Phase} &
\textrm{$a$ (\AA)} &
\textrm{$b$ (\AA)} &
\textrm{$c$ (\AA)} \\
\colrule
$C$2/$c$ & 10.804 & 6.302 & 10.074 \\
$R$3$c$ & 5.680 & 5.680 & 5.680 \\
\label{Table1}
\end{tabular}
\end{ruledtabular}
\end{table}

\section{\label{sec:level3}Results and Discussion}

\begin{figure*}[htbp]
\includegraphics[width=1.0\textwidth]{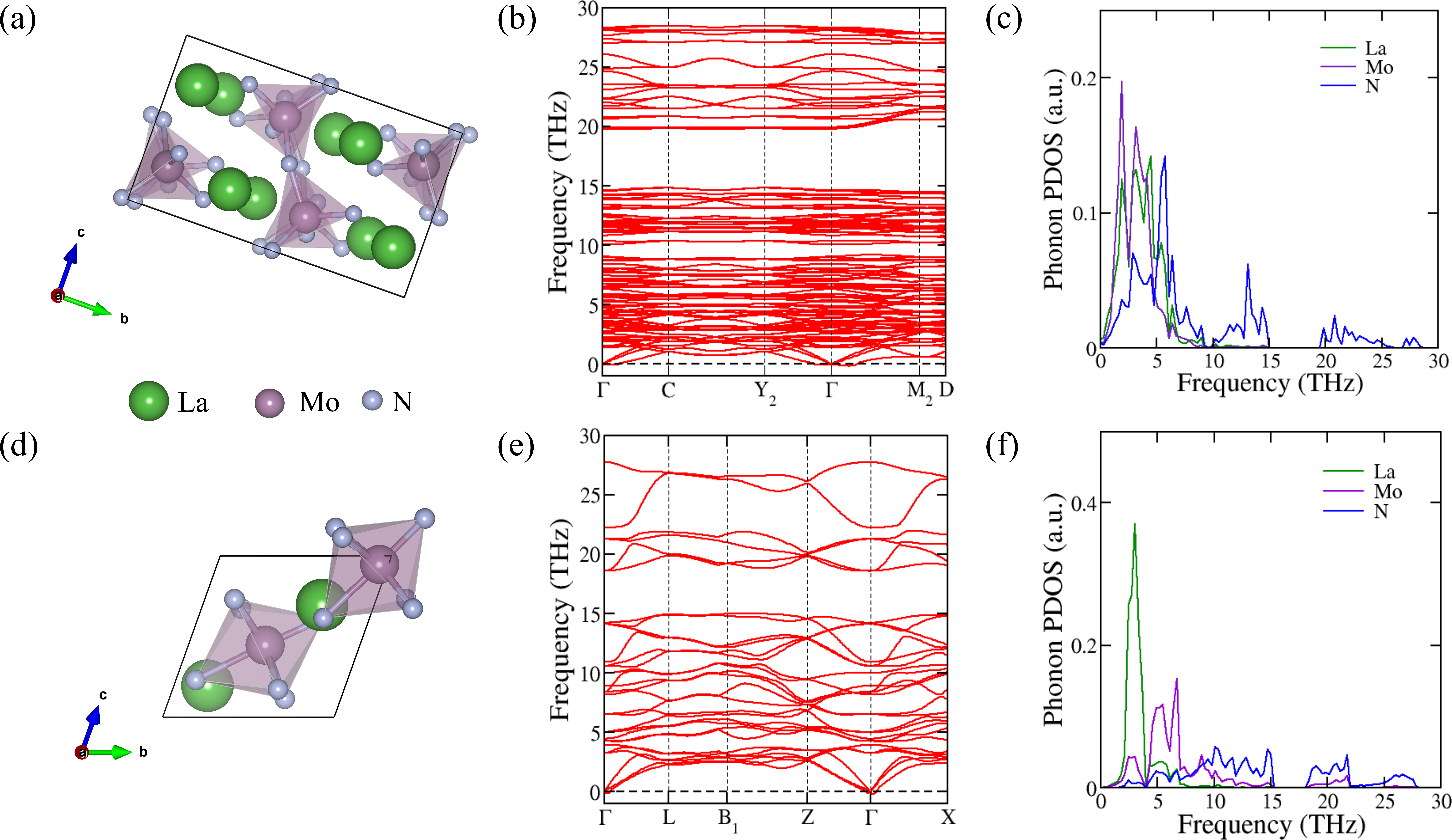}
\caption{Crystal structures (a, d), phonon dispersion (b, e) and projected density of states (PDOS) (c, f) of LaMoN$_3$ in the $C$2/$c$ and $R$3$c$ phase, respectively.}
\label{1}
\end{figure*}

\subsection{Crystal Structure}
The crystal structure of LaMoN$_3$ exists in two main phases: the monoclinic $C$2/$c$ phase~\cite{10.1063/1.5035135, SarmientoPerez2015}, stable under ambient conditions, and the rhombohedral $R$3$c$ phase, which emerges under pressure and exhibits ferroelectricity~\cite{PhysRevB.102.180103}. The $C$2/$c$ phase is centrosymmetric and structurally stable, while the $R$3$c$ phase is non-centrosymmetric, making it useful for ferroelectric applications~\cite{D3SC02171H}. In this work, we have focused our investigation to thoroughly understand the thermal transport properties of the $C$2/$c$ and $R$3$c$ phases of LaMoN$_3$.

The crystal structures of the $C$2/$c$ and $R$3$c$ phases of LaMoN$_3$ are shown in FIG.~\ref{1} (a and d), respectively. The calculated lattice parameters, as listed in TABLE~\ref{Table1}, are consistent with the previously reported values~\cite{doi:10.1021/acs.chemmater.5b02026, PhysRevB.102.180103}.

\subsection{Phonon Dispersion}%
In the $C$2/$c$ and $R$3$c$ phases of LaMoN$_3$, the absence of negative frequencies in the phonon band structure, as shown in FIG.~\ref{1} (b, e), confirms their dynamic stability, ensuring that these phases are mechanically stable. In the $C$2/$c$ phase, the acoustic phonon branches are prominently observed in the low-frequency range of 0-1.85 THz, with a significant contribution from the heavier La and Mo atoms. These acoustic and low-lying optical phonons are strongly coupled, leading to a large projected density of states (PDOS) in the 0-5 THz range (FIG.~\ref{1} (c)) and absence of acoustic-optical gap. Lack of this gap implies a strong interaction between the acoustic and optical phonons, which could potentially enhance phonon-phonon scattering. There is a small optical-optical gap of 0.86 THz in the frequency range of 5-10 THz, and a larger gap of 4.99 THz at higher frequencies. In materials with high phonon scattering rates, such as those with coupled acoustic-optical phonons, the phonon mean free path is shortened, leading to a decrease in thermal conductivity.

In the $R$3$c$ phase of LaMoN$_3$, the acoustic phonon branches are predominantly concentrated in the low-frequency range, extending from 0 to 3.28 THz (see FIG.~\ref{1} (e)). This phase exhibits a small acoustic-optical gap of 0.34 THz, indicating weaker coupling between acoustic and optical phonons than the $C$2/$c$ phase. The optical-optical gap is 3.56 THz. As shown in the PDOS in FIG.~\ref{1} (f), the La atoms primarily contribute to the acoustic phonons, with some contribution from the Mo atoms, while the lighter N atoms dominate the optical phonons. The presence of a small gap in the $R$3$c$ phase suggests that three-phonon scattering in this phase may be less pronounced than in the $C$2/$c$ phase, potentially resulting in slightly higher thermal conductivity. 
\begin{figure*}[htbp]
\includegraphics[width=0.8\textwidth]{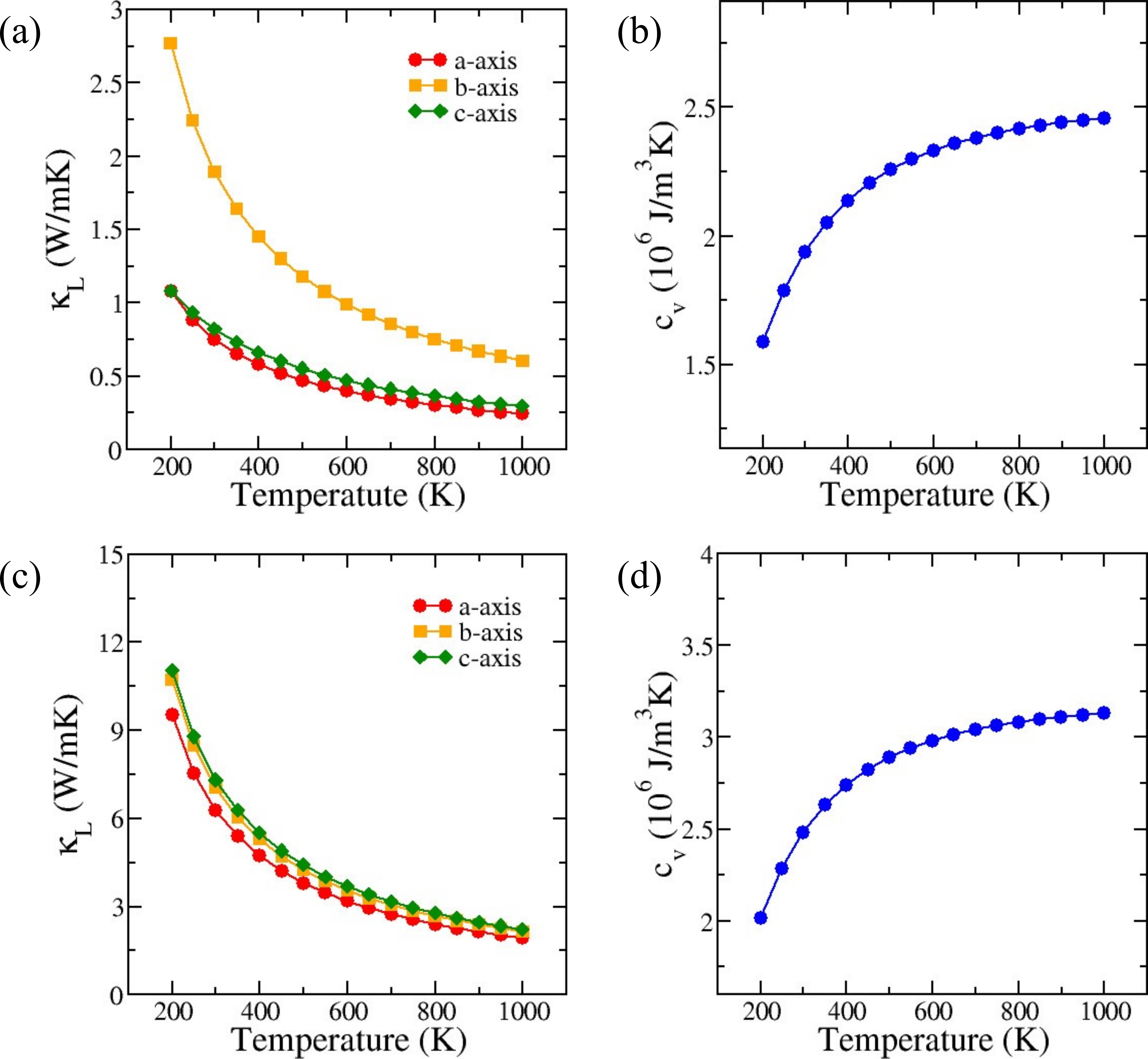}
\caption{Temperature dependence of lattice thermal conductivity (a, c), and specific heat (b, d) of LaMoN$_3$ in the $C$2/$c$ and $R$3$c$ phase, respectively.}
\label{2}
\end{figure*}
\subsection{Lattice Thermal Conductivity}%
By solving the phonon BTE, the computed $\kappa_L$ of LaMoN$_3$ in the $C$2/$c$ and $R$3$c$ phases along the a, b, and c axes as a function of temperature is shown in FIG.~\ref{2} (a, c). The $\kappa_L$ decreases with increasing temperature, approximately following a T$^{-1}$ relationship, as expected for typical materials where phonon scattering becomes more significant at higher temperatures. For the $C$2/$c$ phase of LaMoN$_3$, the values of $\kappa_L$ at 300 K along the a, b, and c directions are 0.75, 1.89, and 0.82 W/mK, respectively. These values indicate that heat transport in the $C$2/$c$ phase is anisotropic along the a and b axes, but exhibits more isotropic behavior along the a and c axes. Specifically, the higher thermal conductivity along the b-axis (1.89 W/mK) suggests that phonon transport is more efficient in this direction, possibly due to more favorable atomic interactions or less phonon scattering. Additionally, the specific heat in the $C$2/$c$ phase, which ranges from 1.58$\times$10$^6$ J/m$^3$K to 2.45$\times$10$^6$ J/m$^3$K, increases with rising temperature, as shown in FIG.~\ref{2}(b). This increase in specific heat is consistent with the general behavior of solids, where the material’s ability to store thermal energy increases as temperature rises.

\begin{figure*}[htbp]
\includegraphics[width=0.8\textwidth]{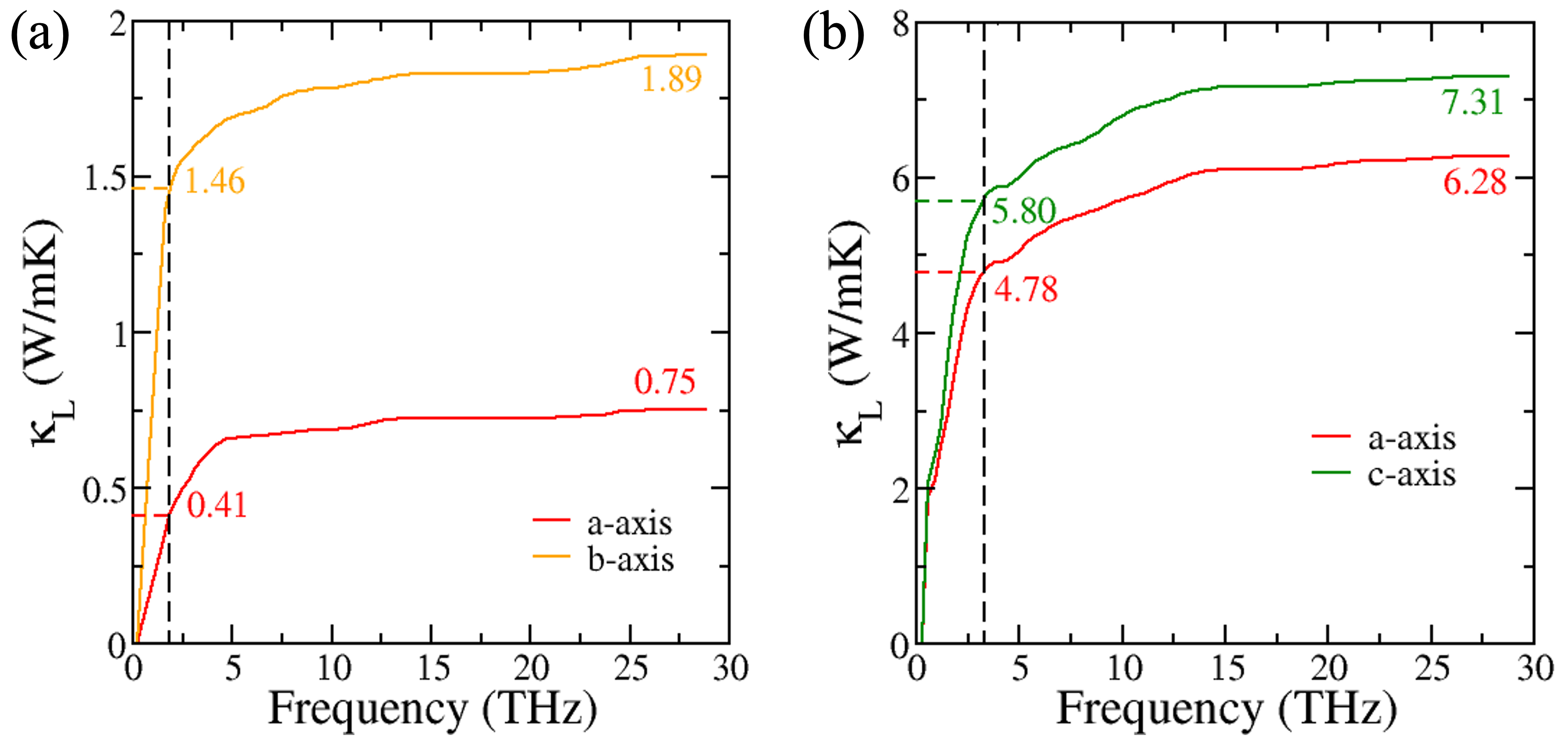}
\caption{Frequency dependence of cumulative $\kappa_L$ at room temperature for LaMoN$_3$ in the (a) $C$2/$c$ and (b) $R$3$c$ phase, respectively. Black dashed line represents the acoustic cut-off.}
\label{3}
\end{figure*}

For the $R$3$c$ phase of LaMoN$_3$, the thermal conductivity values at 300 K along the a, b, and c directions are 6.28, 7.05, and 7.31 W/mK, respectively (see FIG.~\ref{2}(c)). These values are very close to each other, indicating nearly isotropic heat transport in the $R$3$c$ phase. As shown in FIG.~\ref{2} (d), the specific heat in the $R$3$c$ phase increases with temperature, ranging from 2.01$\times$10$^6$ J/m$^3$K to 3.13$\times$10$^6$ J/m$^3$K. This increase further corroborates the trend of higher specific heat with rising temperature and supports the idea that phonon contributions to specific heat and thermal conductivity are coupled~\cite{slack1979thermal, ziman1960electrons}.

In addition to temperature-dependent behavior, the frequency dependence of cumulative $\kappa_L$ at room temperature for both phases provides further insight into the contribution of the phonon to thermal transport. For the $C$2/$c$ phase of LaMoN$_3$, as shown in FIG.~\ref{3}(a), the acoustic phonons make a substantial contribution to the thermal conductivity. The cumulative $\kappa_L$ values for the acoustic modes are 0.41 W/mK and 1.46 W/mK along the a and b axes, respectively. These contributions represent 54\% and 77\% of the total thermal conductivity, highlighting the dominant role of acoustic phonons in heat transport. In the case of the $R$3$c$ phase, the cumulative $\kappa_L$ values along the a and c axes increase to 4.78 W/mK and 5.80 W/mK in the frequency range of 0-3.28 THz (see FIG.~\ref{3}(b)). These values represent 76\% and 79\% of the total thermal conductivity, respectively, again underscoring the dominant role of the acoustic phonons in determining its thermal conductivity. 

Finally, in both phases, the cumulative $\kappa_L$ does not increase significantly at frequencies above 15 THz, indicating that the optical phonons above the optical-optical gap contribute minimally to the thermal conductivity. This suggests that the higher-frequency optical phonons have limited impact on heat transport. The results highlight the crucial role of acoustic phonons in governing the thermal conductivity of LaMoN$_3$ in both the $C$2/$c$ and $R$3$c$ phases, with a more pronounced contribution in the latter due to more isotropic and efficient phonon transport.

\begin{figure*}
\begin{center}
\includegraphics[width=1.05\textwidth]{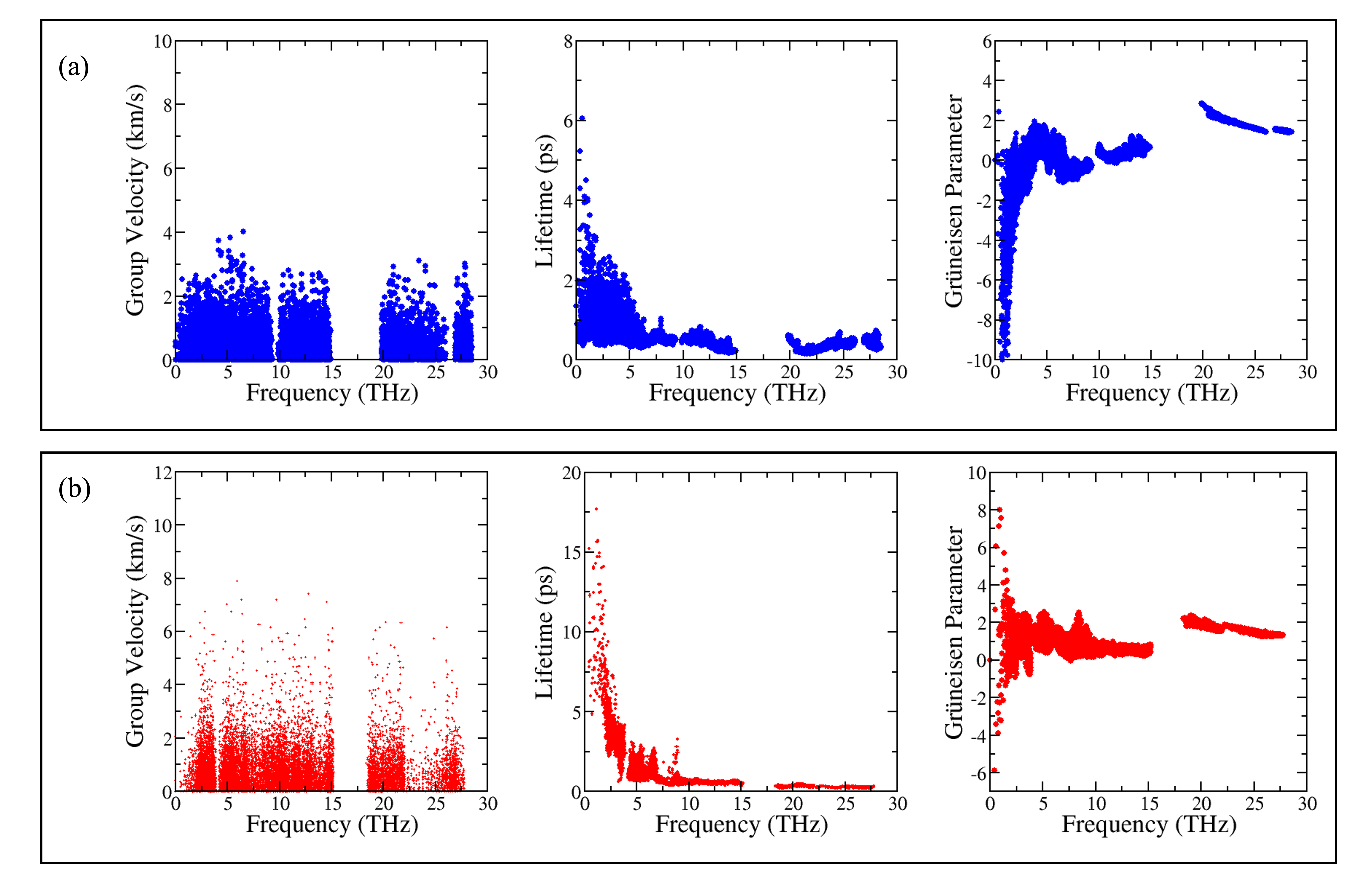}
\caption{Frequency dependence of group velocity, phonon lifetime, and Gr\"uneisen parameter at room temperature in the (a) $C$2/$c$ and (b) $R$3$c$ phase, respectively.}
\label{4}
\end{center}
\end{figure*}

\begin{figure*}[htbp]
\begin{center}
\includegraphics[width=0.8\textwidth]{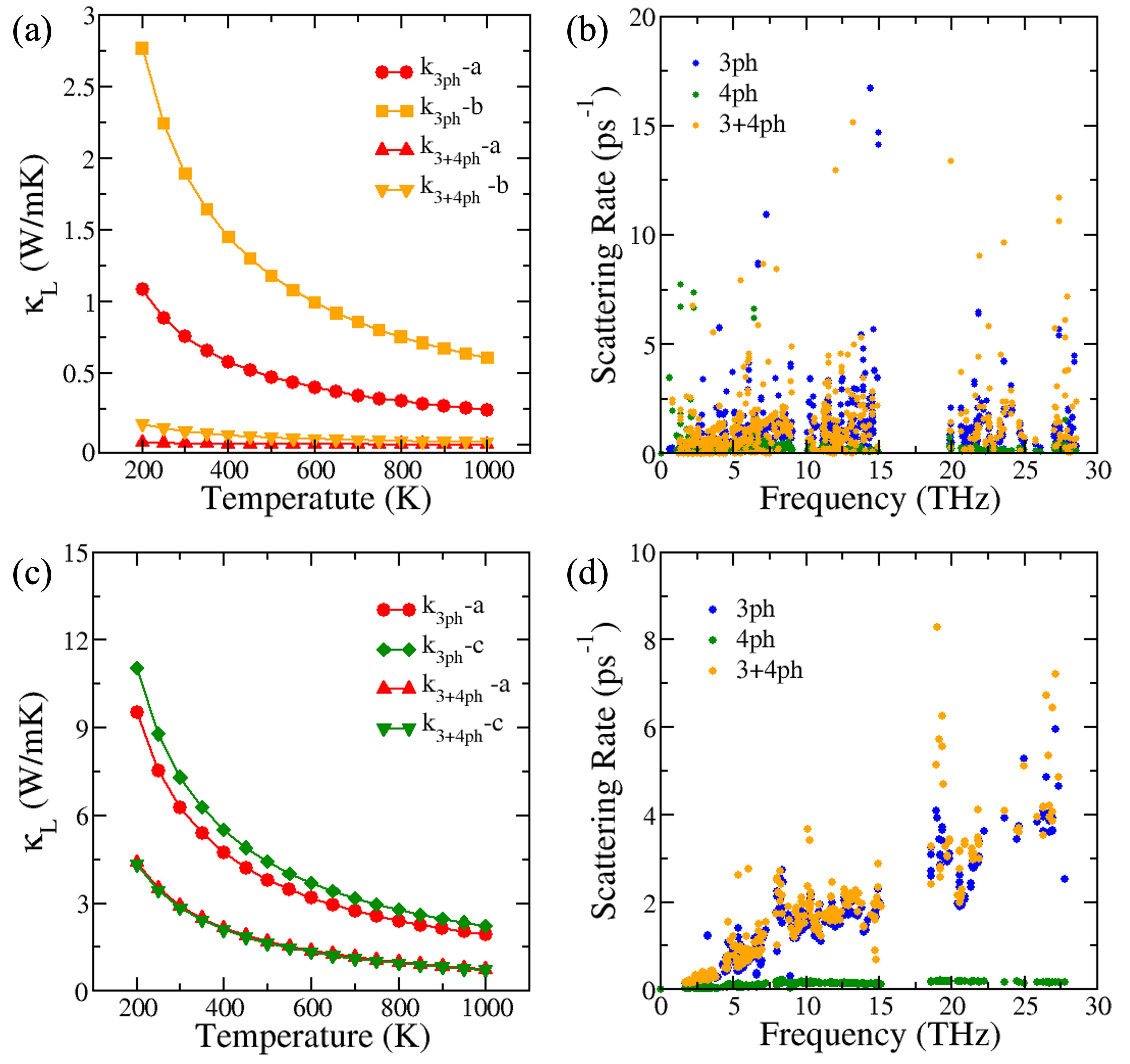}
\caption{Temperature dependence of lattice thermal conductivity (a, c), and scattering rate (b, d) of 3ph, 4ph, and 3 + 4ph processes at room temperature (300K) for LaMoN$_3$ in the $C$2/$c$ and $R$3$c$ phase, respectively.}
\label{5}
\end{center}
\end{figure*} 

\subsection{Group Velocity, Phonon Lifetime and Gr\"uneisen parameter}%
The phonon group velocity and lifetime analysis, as shown in FIG.~\ref{4}(a, b), provides additional insights into the behavior of $\kappa_L$ observed in the $C$2/$c$ and $R$3$c$ phases of LaMoN$_3$. In the low-frequency region below 5 THz, the optical phonons exhibit group velocities comparable to those of acoustic phonons, suggesting that these low-lying optical phonons contribute significantly to heat transport. This observation aligns with the previously discussed absence of an acoustic-optical gap in the $C$2/$c$ phase, which facilitates strong acoustic-optical phonon coupling and enhances phonon-phonon scattering. The increased scattering, in turn, reduces the mean free path of heat-carrying phonons, thereby lowering the overall $\kappa_L$, particularly in the $C$2/$c$ phase, where $\kappa_L$ is found to be much lower than in the $R$3$c$ phase.

However, an interesting aspect arises when examining the group velocities of optical phonons above 5 THz. Notably, these low-frequency optical phonons exhibit group velocities larger than those of the acoustic phonons, suggesting a potential contribution to heat transport. In the $C$2/$c$ phase, these optical phonon modes are primarily associated with the large-scale vibrations of light N atoms, with a smaller but significant contribution from La atoms, as illustrated in FIG.~\ref{1} (c). However, despite their relatively high group velocities, these optical phonons make only a minimal contribution to $\kappa_L$, as indicated by the cumulative thermal conductivity results, which show negligible contributions from phonons above 5 THz. This discrepancy arises from the short lifetimes of these optical modes, which decrease further at higher optical modes, leading to strong phonon scattering and consequently limiting their role in heat transport.

In contrast, in the $R$3$c$ phase, the PDOS (FIG.~\ref{1} (f)) indicates that vibrational modes around 5 THz are mainly contributed by Mo atoms, with small contributions from La and N atoms, while vibrational states around 10 THz are predominantly associated with N atoms. The presence of Mo in low-frequency optical modes may mitigate phonon scattering compared to the $C$2/$c$ phase, where these modes are largely dominated by N atoms. As a result, reduced phonon-phonon scattering in the $R3c$ phase enhances the thermal conductivity of the lattice. Furthermore, despite the large group velocities of optical phonons in both phases, their contribution to $\kappa_L$ remains limited due to their inherently short lifetimes, reinforcing the conclusion that acoustic phonons predominantly govern heat transport in LaMoN$_3$.

Additionally, the Gr\"uneisen parameter ($\gamma$) provides valuable insight into lattice anharmonicity by quantifying changes in phonon frequencies ($\omega$) with respect to variations in unit-cell volume ($V$). Positive and negative values of $\gamma$ indicate phonon frequency softening and hardening due to lattice expansion, respectively, while a large absolute value of $\gamma$ suggests strong lattice anharmonicity. As shown in FIG.~\ref{4}(a, b), phonon modes in the low-frequency range (0–2 THz) exhibit high Gr\"uneisen parameters, ranging from -10 to +4 in the $C$2/$c$ phase and from -6 to +8 in the $R$3$c$ phase of LaMoN$_3$. These large values indicate significant anharmonicity in the low-frequency modes, which correlates with strong phonon-phonon interactions and increased scattering, contributing to the lower $\kappa_L$ in the $C$2/$c$ phase. On the other hand, the high-frequency optical phonon modes possess relatively lower Gr\"uneisen parameters, ranging from -1 to +3 in the $C$2/$c$ phase and 0 to +3 in the $R$3$c$ phase. The lower anharmonicity of high-frequency optical phonons further supports the observation that these modes contribute minimally to thermal conductivity, as they are less susceptible to significant frequency shifts with volume changes and generally have shorter lifetimes.

\subsection{Four Phonon Contributions}%
Four-phonon scattering plays a crucial role in phonon transport, especially in materials with strong anharmonicity and ultralow thermal conductivity \cite{Zeng2023, Yang2019}. In LaMoN$_3$, high Gr\"uneisen parameters observed in low-frequency phonon modes indicate significant lattice anharmonicity, suggesting that four-phonon interactions strongly influence thermal transport. The complex phonon band structure further restricts the satisfaction of three-phonon selection rules, increasing the likelihood of higher-order phonon scattering.

In the $C2/c$ phase of LaMoN$3$, incorporating both three- and four-phonon scattering ($\kappa{3+4ph}$) at 300 K results in an extremely low lattice thermal conductivity of 0.03 W/mK along the a-axis and 0.09 W/mK along the b-axis, marking a drastic 96\% reduction compared to calculations considering only three-phonon scattering (see FIG.\ref{5}(a)). This highlights the dominant role of four-phonon interactions in suppressing thermal transport. The scattering rate analysis in FIG.\ref{5}(b) further confirms this, showing that the four-phonon scattering rate is higher than the three-phonon rate across low-frequency regions, making four-phonon processes the primary contributor to phonon transport in the $C2/c$ phase.

In contrast, the $R{3}c$ phase exhibits a lower sensitivity to four-phonon scattering. At 300 K, incorporating four-phonon interactions reduces the lattice thermal conductivity to 2.83 W/mK along the a-axis and 3.18 W/mK along the c-axis, leading to a 50\% decrease compared to calculations considering only three-phonon scattering. While this suppression is significant, it is notably weaker than in the $C2/c$ phase. As illustrated in FIG.~\ref{5}(c), this reduction persists across the 300–1000 K temperature range, reinforcing that four-phonon interactions still play a crucial role in this phase but are less impactful than in $C2/c$. The comparatively weaker effect can be attributed to the higher phonon group velocities and relatively lower anharmonicity in the $R{3}c$ structure.

Additionally, incorporating four-phonon scattering enhances the relative contribution of optical phonons to $\kappa_L$ in both phases. This effect arises because four-phonon interactions primarily suppress acoustic phonon transport, increasing the role of optical phonons in overall heat conduction. As shown in FIG.~\ref{5}(d), the four-phonon scattering rate is comparable to the three-phonon rate in the low-frequency range, indicating its role in reducing phonon lifetimes. However, in the high-frequency range, particularly in optical phonon branches, the four-phonon scattering rate is lower than the three-phonon rate, suggesting that four-phonon interactions primarily impact acoustic phonons. Consequently, in the $R$3$c$ phase, optical phonons contribute more significantly to thermal conductivity when four-phonon scattering is considered compared to when only three-phonon processes are included.

\section{\label{sec:level4}Conclusions}
In this study, we systematically investigated the lattice thermal conductivity ($\kappa_L$) of LaMoN$_3$ in its $C2/c$ and $R{3}c$ phases using first-principles calculations and the phonon BTE. Our analysis reveals distinct thermal transport behaviors arising from differences in phonon dispersion, group velocity, anharmonicity, and phonon scattering mechanisms.

The C2/c phase exhibits intrinsically low and highly anisotropic thermal conductivity, with $\kappa_L$ values of 0.75, 1.89, and 0.82 W/mK along the a, b, and c axes at 300 K, respectively. The absence of an acoustic-optical gap enhances three-phonon scattering, leading to significantly reduced $\kappa_L$. Similar to conventional low-$\kappa_L$ materials, where four-phonon scattering plays a crucial role, our calculations reveal that this mechanism is also predominant in the C2/c phase of LaMoN$_3$. The inclusion of four-phonon scattering results in an extremely large reduction in $\kappa_L$ of approximately 96\%, underscoring its dominant influence on thermal transport in this phase.

In contrast, the $R3c$ phase exhibits significantly higher and nearly isotropic thermal conductivity, with $\kappa_L$ values of 6.28, 7.05, and 7.31 W/mK along the a, b, and c directions, respectively. The phonon dispersion analysis shows that this phase hosts high-velocity optical phonons, which contribute substantially to heat transport. While anharmonicity is still present, its impact is comparatively lower than in the $C2/c$ phase. Four-phonon scattering reduces $\kappa_L$ by approximately 50\%, playing a notable but less dominant role than in the $C2/c$ phase. The stronger phonon group velocities and reduced phonon-phonon scattering in $R3c$ allow for higher thermal conductivity, with optical phonons contributing more prominently due to their limited four-phonon scattering rates.

Overall, our findings highlight the critical role of phonon-phonon interactions in governing the thermal transport properties of LaMoN$_3$. While the $C2/c$ phase demonstrates ultralow $\kappa_L$ due to strong four-phonon scattering, the $R$3$c$ phase exhibits significantly higher thermal conductivity. These insights provide a deeper understanding of phonon transport in LaMoN$_3$ and offer valuable guidance for designing nitride perovskites with tailored thermal properties for thermoelectric and thermal management applications.

\section{\label{sec:level5}ACKNOWLEDGMENTS}
M.J. acknowledges CSIR, India, for the senior research fellowship (Grant [09/086(1344)/2018-EMR-I]). S.M. acknowledges IIT Delhi for the senior research fellowship. S.B. acknowledges financial support from SERB under a core research grant (Grant CRG/2019/000647) to set up his High Performance Computing (HPC) facility “Veena” at IIT Delhi for computational resources.
\nocite{*}

\bibliography{references}% Produces the bibliography via BibTeX.

\begin{thebibliography}{47}%
\makeatletter
\providecommand \@ifxundefined [1]{%
 \@ifx{#1\undefined}
}%
\providecommand \@ifnum [1]{%
 \ifnum #1\expandafter \@firstoftwo
 \else \expandafter \@secondoftwo
 \fi
}%
\providecommand \@ifx [1]{%
 \ifx #1\expandafter \@firstoftwo
 \else \expandafter \@secondoftwo
 \fi
}%
\providecommand \natexlab [1]{#1}%
\providecommand \enquote  [1]{``#1''}%
\providecommand \bibnamefont  [1]{#1}%
\providecommand \bibfnamefont [1]{#1}%
\providecommand \citenamefont [1]{#1}%
\providecommand \href@noop [0]{\@secondoftwo}%
\providecommand \href [0]{\begingroup \@sanitize@url \@href}%
\providecommand \@href[1]{\@@startlink{#1}\@@href}%
\providecommand \@@href[1]{\endgroup#1\@@endlink}%
\providecommand \@sanitize@url [0]{\catcode `\\12\catcode `\$12\catcode `\&12\catcode `\#12\catcode `\^12\catcode `\_12\catcode `\%12\relax}%
\providecommand \@@startlink[1]{}%
\providecommand \@@endlink[0]{}%
\providecommand \url  [0]{\begingroup\@sanitize@url \@url }%
\providecommand \@url [1]{\endgroup\@href {#1}{\urlprefix }}%
\providecommand \urlprefix  [0]{URL }%
\providecommand \Eprint [0]{\href }%
\providecommand \doibase [0]{https://doi.org/}%
\providecommand \selectlanguage [0]{\@gobble}%
\providecommand \bibinfo  [0]{\@secondoftwo}%
\providecommand \bibfield  [0]{\@secondoftwo}%
\providecommand \translation [1]{[#1]}%
\providecommand \BibitemOpen [0]{}%
\providecommand \bibitemStop [0]{}%
\providecommand \bibitemNoStop [0]{.\EOS\space}%
\providecommand \EOS [0]{\spacefactor3000\relax}%
\providecommand \BibitemShut  [1]{\csname bibitem#1\endcsname}%
\let\auto@bib@innerbib\@empty
%</preamble>
\bibitem [{\citenamefont {Shi}\ \emph {et~al.}(2020)\citenamefont {Shi}, \citenamefont {Yu}, \citenamefont {Wang}, \citenamefont {Ye}, \citenamefont {Gong}, \citenamefont {Ma}, \citenamefont {Jiang}, \citenamefont {Hua}, \citenamefont {Shuai}, \citenamefont {Zhang},\ and\ \citenamefont {Ye}}]{Shi2020}%
  \BibitemOpen
  \bibfield  {author} {\bibinfo {author} {\bibfnamefont {C.}~\bibnamefont {Shi}}, \bibinfo {author} {\bibfnamefont {H.}~\bibnamefont {Yu}}, \bibinfo {author} {\bibfnamefont {Q.~W.}\ \bibnamefont {Wang}}, \bibinfo {author} {\bibfnamefont {L.}~\bibnamefont {Ye}}, \bibinfo {author} {\bibfnamefont {Z.~X.}\ \bibnamefont {Gong}}, \bibinfo {author} {\bibfnamefont {J.~J.}\ \bibnamefont {Ma}}, \bibinfo {author} {\bibfnamefont {J.~Y.}\ \bibnamefont {Jiang}}, \bibinfo {author} {\bibfnamefont {M.~M.}\ \bibnamefont {Hua}}, \bibinfo {author} {\bibfnamefont {C.}~\bibnamefont {Shuai}}, \bibinfo {author} {\bibfnamefont {Y.}~\bibnamefont {Zhang}},\ and\ \bibinfo {author} {\bibfnamefont {H.~Y.}\ \bibnamefont {Ye}},\ }\bibfield  {title} {\bibinfo {title} {Hybrid organic-inorganic antiperovskites},\ }\href@noop {} {\bibfield  {journal} {\bibinfo  {journal} {Angew. Chem. Int. Ed.}\ }\textbf {\bibinfo {volume} {59}},\ \bibinfo {pages} {167} (\bibinfo {year} {2020})}\BibitemShut {NoStop}%
\bibitem [{\citenamefont {Jain}\ \emph {et~al.}(2020)\citenamefont {Jain}, \citenamefont {Singh}, \citenamefont {Basera}, \citenamefont {Kumar},\ and\ \citenamefont {Bhattacharya}}]{D0TC01484B}%
  \BibitemOpen
  \bibfield  {author} {\bibinfo {author} {\bibfnamefont {M.}~\bibnamefont {Jain}}, \bibinfo {author} {\bibfnamefont {A.}~\bibnamefont {Singh}}, \bibinfo {author} {\bibfnamefont {P.}~\bibnamefont {Basera}}, \bibinfo {author} {\bibfnamefont {M.}~\bibnamefont {Kumar}},\ and\ \bibinfo {author} {\bibfnamefont {S.}~\bibnamefont {Bhattacharya}},\ }\bibfield  {title} {\bibinfo {title} {Understanding the role of {Sn} substitution and {Pb}-vac in enhancing the optical properties and solar cell efficiency of {CH(NH$_2$)$_2$Pb$_{1-x-y}$Sn$_x$vac$_y$Br$_3$}},\ }\href@noop {} {\bibfield  {journal} {\bibinfo  {journal} {J. Mater. Chem. C}\ }\textbf {\bibinfo {volume} {8}},\ \bibinfo {pages} {10362} (\bibinfo {year} {2020})}\BibitemShut {NoStop}%
\bibitem [{\citenamefont {Lee}\ \emph {et~al.}(2015)\citenamefont {Lee}, \citenamefont {Lu}, \citenamefont {Gu}, \citenamefont {Choi}, \citenamefont {Li}, \citenamefont {Ryu}, \citenamefont {Paudel}, \citenamefont {Song}, \citenamefont {Mikheev},\ and\ \citenamefont {et~al.}}]{Lee2015}%
  \BibitemOpen
  \bibfield  {author} {\bibinfo {author} {\bibfnamefont {D.}~\bibnamefont {Lee}}, \bibinfo {author} {\bibfnamefont {H.}~\bibnamefont {Lu}}, \bibinfo {author} {\bibfnamefont {Y.}~\bibnamefont {Gu}}, \bibinfo {author} {\bibfnamefont {S.~Y.}\ \bibnamefont {Choi}}, \bibinfo {author} {\bibfnamefont {S.~D.}\ \bibnamefont {Li}}, \bibinfo {author} {\bibfnamefont {S.}~\bibnamefont {Ryu}}, \bibinfo {author} {\bibfnamefont {T.~R.}\ \bibnamefont {Paudel}}, \bibinfo {author} {\bibfnamefont {K.}~\bibnamefont {Song}}, \bibinfo {author} {\bibfnamefont {E.}~\bibnamefont {Mikheev}},\ and\ \bibinfo {author} {\bibfnamefont {S.~L.}\ \bibnamefont {et~al.}},\ }\bibfield  {title} {\bibinfo {title} {Emergence of room-temperature ferroelectricity at reduced dimensions},\ }\href@noop {} {\bibfield  {journal} {\bibinfo  {journal} {Science}\ }\textbf {\bibinfo {volume} {349}},\ \bibinfo {pages} {1314} (\bibinfo {year} {2015})}\BibitemShut {NoStop}%
\bibitem [{\citenamefont {Tang}\ \emph {et~al.}(2015)\citenamefont {Tang}, \citenamefont {Zhu}, \citenamefont {Ma}, \citenamefont {Borisevich}, \citenamefont {Morozovska}, \citenamefont {Eliseev}, \citenamefont {Wang}, \citenamefont {Wang}, \citenamefont {Xu}, \citenamefont {Zhang},\ and\ \citenamefont {Pennycook}}]{Tang2015}%
  \BibitemOpen
  \bibfield  {author} {\bibinfo {author} {\bibfnamefont {Y.~L.}\ \bibnamefont {Tang}}, \bibinfo {author} {\bibfnamefont {Y.~L.}\ \bibnamefont {Zhu}}, \bibinfo {author} {\bibfnamefont {X.~L.}\ \bibnamefont {Ma}}, \bibinfo {author} {\bibfnamefont {A.~Y.}\ \bibnamefont {Borisevich}}, \bibinfo {author} {\bibfnamefont {A.~N.}\ \bibnamefont {Morozovska}}, \bibinfo {author} {\bibfnamefont {E.~A.}\ \bibnamefont {Eliseev}}, \bibinfo {author} {\bibfnamefont {W.~Y.}\ \bibnamefont {Wang}}, \bibinfo {author} {\bibfnamefont {Y.~J.}\ \bibnamefont {Wang}}, \bibinfo {author} {\bibfnamefont {Y.~B.}\ \bibnamefont {Xu}}, \bibinfo {author} {\bibfnamefont {Z.~D.}\ \bibnamefont {Zhang}},\ and\ \bibinfo {author} {\bibfnamefont {S.~J.}\ \bibnamefont {Pennycook}},\ }\bibfield  {title} {\bibinfo {title} {Observation of a periodic array of flux-closure quadrants in strained ferroelectric {PbTiO$_3$} films},\ }\href@noop {} {\bibfield  {journal} {\bibinfo  {journal} {Science}\ }\textbf {\bibinfo {volume} {348}},\ \bibinfo {pages} {547}
  (\bibinfo {year} {2015})}\BibitemShut {NoStop}%
\bibitem [{\citenamefont {Kojima}\ \emph {et~al.}(2009)\citenamefont {Kojima}, \citenamefont {Teshima}, \citenamefont {Shirai},\ and\ \citenamefont {Miyasaka}}]{Kojima2009}%
  \BibitemOpen
  \bibfield  {author} {\bibinfo {author} {\bibfnamefont {A.}~\bibnamefont {Kojima}}, \bibinfo {author} {\bibfnamefont {K.}~\bibnamefont {Teshima}}, \bibinfo {author} {\bibfnamefont {Y.}~\bibnamefont {Shirai}},\ and\ \bibinfo {author} {\bibfnamefont {T.}~\bibnamefont {Miyasaka}},\ }\bibfield  {title} {\bibinfo {title} {Organometal halide perovskites as visible-light sensitizers for photovoltaic cells},\ }\href@noop {} {\bibfield  {journal} {\bibinfo  {journal} {J. Am. Chem. Soc.}\ }\textbf {\bibinfo {volume} {131}},\ \bibinfo {pages} {6050} (\bibinfo {year} {2009})}\BibitemShut {NoStop}%
\bibitem [{\citenamefont {Dimos}\ and\ \citenamefont {Mueller}(1998)}]{Dimos1998}%
  \BibitemOpen
  \bibfield  {author} {\bibinfo {author} {\bibfnamefont {D.}~\bibnamefont {Dimos}}\ and\ \bibinfo {author} {\bibfnamefont {C.~H.}\ \bibnamefont {Mueller}},\ }\bibfield  {title} {\bibinfo {title} {Perovskite thin films for high-frequency capacitor applications: Metal oxides},\ }\href@noop {} {\bibfield  {journal} {\bibinfo  {journal} {Annu. Rev. Mater. Sci}\ }\textbf {\bibinfo {volume} {28}},\ \bibinfo {pages} {397} (\bibinfo {year} {1998})}\BibitemShut {NoStop}%
\bibitem [{\citenamefont {Muralt}\ \emph {et~al.}(2009)\citenamefont {Muralt}, \citenamefont {Polcawich},\ and\ \citenamefont {Trolier-McKinstry}}]{Muralt2009}%
  \BibitemOpen
  \bibfield  {author} {\bibinfo {author} {\bibfnamefont {P.}~\bibnamefont {Muralt}}, \bibinfo {author} {\bibfnamefont {R.~G.}\ \bibnamefont {Polcawich}},\ and\ \bibinfo {author} {\bibfnamefont {S.}~\bibnamefont {Trolier-McKinstry}},\ }\bibfield  {title} {\bibinfo {title} {Piezoelectric thin films for sensors, actuators, and energy harvesting},\ }\href@noop {} {\bibfield  {journal} {\bibinfo  {journal} {MRS Bull.}\ }\textbf {\bibinfo {volume} {34}},\ \bibinfo {pages} {658} (\bibinfo {year} {2009})}\BibitemShut {NoStop}%
\bibitem [{\citenamefont {Papac}\ \emph {et~al.}(2021)\citenamefont {Papac}, \citenamefont {Stevanovi{\'c}}, \citenamefont {Zakutayev},\ and\ \citenamefont {O'Hayre}}]{Papac2021}%
  \BibitemOpen
  \bibfield  {author} {\bibinfo {author} {\bibfnamefont {M.}~\bibnamefont {Papac}}, \bibinfo {author} {\bibfnamefont {V.}~\bibnamefont {Stevanovi{\'c}}}, \bibinfo {author} {\bibfnamefont {A.}~\bibnamefont {Zakutayev}},\ and\ \bibinfo {author} {\bibfnamefont {R.}~\bibnamefont {O'Hayre}},\ }\bibfield  {title} {\bibinfo {title} {Triple ionic–electronic conducting oxides for next-generation electrochemical devices},\ }\href@noop {} {\bibfield  {journal} {\bibinfo  {journal} {Nat. Mater.}\ }\textbf {\bibinfo {volume} {20}},\ \bibinfo {pages} {301} (\bibinfo {year} {2021})}\BibitemShut {NoStop}%
\bibitem [{\citenamefont {Boris}\ \emph {et~al.}(2011)\citenamefont {Boris}, \citenamefont {Matiks}, \citenamefont {Benckiser}, \citenamefont {Frano}, \citenamefont {Popovich}, \citenamefont {Hinkov}, \citenamefont {Wochner}, \citenamefont {Castro-Colin}, \citenamefont {Detemple},\ and\ \citenamefont {et~al.}}]{Boris2011}%
  \BibitemOpen
  \bibfield  {author} {\bibinfo {author} {\bibfnamefont {A.~V.}\ \bibnamefont {Boris}}, \bibinfo {author} {\bibfnamefont {Y.}~\bibnamefont {Matiks}}, \bibinfo {author} {\bibfnamefont {E.}~\bibnamefont {Benckiser}}, \bibinfo {author} {\bibfnamefont {A.}~\bibnamefont {Frano}}, \bibinfo {author} {\bibfnamefont {P.}~\bibnamefont {Popovich}}, \bibinfo {author} {\bibfnamefont {V.}~\bibnamefont {Hinkov}}, \bibinfo {author} {\bibfnamefont {P.}~\bibnamefont {Wochner}}, \bibinfo {author} {\bibfnamefont {M.}~\bibnamefont {Castro-Colin}}, \bibinfo {author} {\bibfnamefont {E.}~\bibnamefont {Detemple}},\ and\ \bibinfo {author} {\bibfnamefont {V.~K.~M.}\ \bibnamefont {et~al.}},\ }\bibfield  {title} {\bibinfo {title} {Dimensionality control of electronic phase transitions in nickel-oxide superlattices},\ }\href@noop {} {\bibfield  {journal} {\bibinfo  {journal} {Science}\ }\textbf {\bibinfo {volume} {332}},\ \bibinfo {pages} {937} (\bibinfo {year} {2011})}\BibitemShut {NoStop}%
\bibitem [{\citenamefont {Kim}\ \emph {et~al.}(2020)\citenamefont {Kim}, \citenamefont {Lee}, \citenamefont {Jung}, \citenamefont {Shin},\ and\ \citenamefont {Park}}]{Kim2020}%
  \BibitemOpen
  \bibfield  {author} {\bibinfo {author} {\bibfnamefont {J.~Y.}\ \bibnamefont {Kim}}, \bibinfo {author} {\bibfnamefont {J.-W.}\ \bibnamefont {Lee}}, \bibinfo {author} {\bibfnamefont {H.~S.}\ \bibnamefont {Jung}}, \bibinfo {author} {\bibfnamefont {H.}~\bibnamefont {Shin}},\ and\ \bibinfo {author} {\bibfnamefont {N.-G.}\ \bibnamefont {Park}},\ }\bibfield  {title} {\bibinfo {title} {High-efficiency perovskite solar cells},\ }\href@noop {} {\bibfield  {journal} {\bibinfo  {journal} {Chem. Rev.}\ }\textbf {\bibinfo {volume} {120}},\ \bibinfo {pages} {7867} (\bibinfo {year} {2020})}\BibitemShut {NoStop}%
\bibitem [{\citenamefont {Tiwari}\ \emph {et~al.}(2021)\citenamefont {Tiwari}, \citenamefont {Satpute}, \citenamefont {Mehare},\ and\ \citenamefont {Dhoble}}]{Tiwari2021}%
  \BibitemOpen
  \bibfield  {author} {\bibinfo {author} {\bibfnamefont {A.}~\bibnamefont {Tiwari}}, \bibinfo {author} {\bibfnamefont {N.~S.}\ \bibnamefont {Satpute}}, \bibinfo {author} {\bibfnamefont {C.~M.}\ \bibnamefont {Mehare}},\ and\ \bibinfo {author} {\bibfnamefont {S.~J.}\ \bibnamefont {Dhoble}},\ }\bibfield  {title} {\bibinfo {title} {Challenges, recent advances and improvements for enhancing the efficiencies of {ABX$_3$}-based peleds (perovskites light emitting diodes): A review},\ }\href@noop {} {\bibfield  {journal} {\bibinfo  {journal} {J. Alloys Compd.}\ }\textbf {\bibinfo {volume} {850}},\ \bibinfo {pages} {156827} (\bibinfo {year} {2021})}\BibitemShut {NoStop}%
\bibitem [{\citenamefont {Gill}\ \emph {et~al.}(2024)\citenamefont {Gill}, \citenamefont {Jain}, \citenamefont {Bhumla}, \citenamefont {Basera}, \citenamefont {Kumar},\ and\ \citenamefont {Bhattacharya}}]{Gill2024}%
  \BibitemOpen
  \bibfield  {author} {\bibinfo {author} {\bibfnamefont {D.}~\bibnamefont {Gill}}, \bibinfo {author} {\bibfnamefont {M.}~\bibnamefont {Jain}}, \bibinfo {author} {\bibfnamefont {P.}~\bibnamefont {Bhumla}}, \bibinfo {author} {\bibfnamefont {P.}~\bibnamefont {Basera}}, \bibinfo {author} {\bibfnamefont {M.}~\bibnamefont {Kumar}},\ and\ \bibinfo {author} {\bibfnamefont {S.}~\bibnamefont {Bhattacharya}},\ }\bibinfo {title} {Theoretical insights of designing perovskite materials for optoelectronic applications},\ in\ \href@noop {} {\emph {\bibinfo {booktitle} {Perovskite Optoelectronic Devices}}},\ \bibinfo {editor} {edited by\ \bibinfo {editor} {\bibfnamefont {B.}~\bibnamefont {Pradhan}}}\ (\bibinfo  {publisher} {Springer International Publishing},\ \bibinfo {address} {Cham},\ \bibinfo {year} {2024})\ pp.\ \bibinfo {pages} {113--148}\BibitemShut {NoStop}%
\bibitem [{\citenamefont {Wang}\ and\ \citenamefont {Kim}(2017)}]{Wang2017}%
  \BibitemOpen
  \bibfield  {author} {\bibinfo {author} {\bibfnamefont {H.}~\bibnamefont {Wang}}\ and\ \bibinfo {author} {\bibfnamefont {D.~H.}\ \bibnamefont {Kim}},\ }\bibfield  {title} {\bibinfo {title} {Perovskite-based photodetectors: Materials and devices},\ }\href@noop {} {\bibfield  {journal} {\bibinfo  {journal} {Chem. Soc. Rev.}\ }\textbf {\bibinfo {volume} {46}},\ \bibinfo {pages} {5204} (\bibinfo {year} {2017})}\BibitemShut {NoStop}%
\bibitem [{\citenamefont {Zakutayev}(2016)}]{Zakutayev2016}%
  \BibitemOpen
  \bibfield  {author} {\bibinfo {author} {\bibfnamefont {A.}~\bibnamefont {Zakutayev}},\ }\bibfield  {title} {\bibinfo {title} {Design of nitride semiconductors for solar energy conversion},\ }\href@noop {} {\bibfield  {journal} {\bibinfo  {journal} {J. Mater. Chem. A}\ }\textbf {\bibinfo {volume} {4}},\ \bibinfo {pages} {6742} (\bibinfo {year} {2016})}\BibitemShut {NoStop}%
\bibitem [{\citenamefont {Jain}\ \emph {et~al.}(2023)\citenamefont {Jain}, \citenamefont {Gill}, \citenamefont {Monga},\ and\ \citenamefont {Bhattacharya}}]{Jain2023Oxynitride}%
  \BibitemOpen
  \bibfield  {author} {\bibinfo {author} {\bibfnamefont {M.}~\bibnamefont {Jain}}, \bibinfo {author} {\bibfnamefont {D.}~\bibnamefont {Gill}}, \bibinfo {author} {\bibfnamefont {S.}~\bibnamefont {Monga}},\ and\ \bibinfo {author} {\bibfnamefont {S.}~\bibnamefont {Bhattacharya}},\ }\bibfield  {title} {\bibinfo {title} {Oxynitride, oxyfluoride, and nitrofluoride perovskites: Theoretical evaluation of photon absorption properties for solar water splitting},\ }\href {https://doi.org/10.1021/acs.jpcc.3c03449} {\bibfield  {journal} {\bibinfo  {journal} {J. Phys. Chem. C}\ }\textbf {\bibinfo {volume} {127}},\ \bibinfo {pages} {15620} (\bibinfo {year} {2023})}\BibitemShut {NoStop}%
\bibitem [{\citenamefont {Sarmiento-Pérez}\ \emph {et~al.}(2015)\citenamefont {Sarmiento-Pérez}, \citenamefont {Cerqueira}, \citenamefont {Körbel}, \citenamefont {Botti},\ and\ \citenamefont {Marques}}]{SarmientoPerez2015}%
  \BibitemOpen
  \bibfield  {author} {\bibinfo {author} {\bibfnamefont {R.}~\bibnamefont {Sarmiento-Pérez}}, \bibinfo {author} {\bibfnamefont {T.~F.~T.}\ \bibnamefont {Cerqueira}}, \bibinfo {author} {\bibfnamefont {S.}~\bibnamefont {Körbel}}, \bibinfo {author} {\bibfnamefont {S.}~\bibnamefont {Botti}},\ and\ \bibinfo {author} {\bibfnamefont {M.~A.~L.}\ \bibnamefont {Marques}},\ }\bibfield  {title} {\bibinfo {title} {Prediction of stable nitride perovskites},\ }\href@noop {} {\bibfield  {journal} {\bibinfo  {journal} {Chem. Mater.}\ }\textbf {\bibinfo {volume} {27}},\ \bibinfo {pages} {5957} (\bibinfo {year} {2015})}\BibitemShut {NoStop}%
\bibitem [{\citenamefont {Talley}\ \emph {et~al.}(2021)\citenamefont {Talley}, \citenamefont {Perkins}, \citenamefont {Diercks}, \citenamefont {Brennecka},\ and\ \citenamefont {Zakutayev}}]{Talley2021}%
  \BibitemOpen
  \bibfield  {author} {\bibinfo {author} {\bibfnamefont {K.~R.}\ \bibnamefont {Talley}}, \bibinfo {author} {\bibfnamefont {C.~L.}\ \bibnamefont {Perkins}}, \bibinfo {author} {\bibfnamefont {D.~R.}\ \bibnamefont {Diercks}}, \bibinfo {author} {\bibfnamefont {G.~L.}\ \bibnamefont {Brennecka}},\ and\ \bibinfo {author} {\bibfnamefont {A.}~\bibnamefont {Zakutayev}},\ }\bibfield  {title} {\bibinfo {title} {Synthesis of {LaWN$_3$} nitride perovskite with polar symmetry},\ }\href@noop {} {\bibfield  {journal} {\bibinfo  {journal} {Science}\ }\textbf {\bibinfo {volume} {374}},\ \bibinfo {pages} {1488} (\bibinfo {year} {2021})}\BibitemShut {NoStop}%
\bibitem [{\citenamefont {Sherbondy}\ \emph {et~al.}(2022)\citenamefont {Sherbondy}, \citenamefont {Smaha}, \citenamefont {Bartel}, \citenamefont {Holtz}, \citenamefont {Talley}, \citenamefont {Levy-Wendt}, \citenamefont {Perkins}, \citenamefont {Eley}, \citenamefont {Zakutayev},\ and\ \citenamefont {Brennecka}}]{Sherbondy2022}%
  \BibitemOpen
  \bibfield  {author} {\bibinfo {author} {\bibfnamefont {R.}~\bibnamefont {Sherbondy}}, \bibinfo {author} {\bibfnamefont {R.~W.}\ \bibnamefont {Smaha}}, \bibinfo {author} {\bibfnamefont {C.~J.}\ \bibnamefont {Bartel}}, \bibinfo {author} {\bibfnamefont {M.~E.}\ \bibnamefont {Holtz}}, \bibinfo {author} {\bibfnamefont {K.~R.}\ \bibnamefont {Talley}}, \bibinfo {author} {\bibfnamefont {B.}~\bibnamefont {Levy-Wendt}}, \bibinfo {author} {\bibfnamefont {C.~L.}\ \bibnamefont {Perkins}}, \bibinfo {author} {\bibfnamefont {S.}~\bibnamefont {Eley}}, \bibinfo {author} {\bibfnamefont {A.}~\bibnamefont {Zakutayev}},\ and\ \bibinfo {author} {\bibfnamefont {G.~L.}\ \bibnamefont {Brennecka}},\ }\bibfield  {title} {\bibinfo {title} {High-throughput selection and experimental realization of two new {Ce}-based nitride perovskites: {CeMoN$_3$} and {CeWN$_3$}},\ }\href@noop {} {\bibfield  {journal} {\bibinfo  {journal} {Chem. Mater.}\ }\textbf {\bibinfo {volume} {34}},\ \bibinfo {pages} {6883} (\bibinfo {year} {2022})}\BibitemShut
  {NoStop}%
\bibitem [{\citenamefont {Fang}\ \emph {et~al.}(2017)\citenamefont {Fang}, \citenamefont {Fisher}, \citenamefont {Kuwabara}, \citenamefont {Shen}, \citenamefont {Ogawa}, \citenamefont {Moriwake}, \citenamefont {Huang},\ and\ \citenamefont {Duan}}]{PhysRevB.95.014111}%
  \BibitemOpen
  \bibfield  {author} {\bibinfo {author} {\bibfnamefont {Y.-W.}\ \bibnamefont {Fang}}, \bibinfo {author} {\bibfnamefont {C.~A.~J.}\ \bibnamefont {Fisher}}, \bibinfo {author} {\bibfnamefont {A.}~\bibnamefont {Kuwabara}}, \bibinfo {author} {\bibfnamefont {X.-W.}\ \bibnamefont {Shen}}, \bibinfo {author} {\bibfnamefont {T.}~\bibnamefont {Ogawa}}, \bibinfo {author} {\bibfnamefont {H.}~\bibnamefont {Moriwake}}, \bibinfo {author} {\bibfnamefont {R.}~\bibnamefont {Huang}},\ and\ \bibinfo {author} {\bibfnamefont {C.-G.}\ \bibnamefont {Duan}},\ }\bibfield  {title} {\bibinfo {title} {Lattice dynamics and ferroelectric properties of the nitride perovskite {LaWN$_{3}$}},\ }\href@noop {} {\bibfield  {journal} {\bibinfo  {journal} {Phys. Rev. B}\ }\textbf {\bibinfo {volume} {95}},\ \bibinfo {pages} {014111} (\bibinfo {year} {2017})}\BibitemShut {NoStop}%
\bibitem [{\citenamefont {Sarmiento-P{\'e}rez}\ \emph {et~al.}(2015)\citenamefont {Sarmiento-P{\'e}rez}, \citenamefont {Cerqueira}, \citenamefont {K{\"o}rbel}, \citenamefont {Botti},\ and\ \citenamefont {Marques}}]{doi:10.1021/acs.chemmater.5b02026}%
  \BibitemOpen
  \bibfield  {author} {\bibinfo {author} {\bibfnamefont {R.}~\bibnamefont {Sarmiento-P{\'e}rez}}, \bibinfo {author} {\bibfnamefont {T.~F.~T.}\ \bibnamefont {Cerqueira}}, \bibinfo {author} {\bibfnamefont {S.}~\bibnamefont {K{\"o}rbel}}, \bibinfo {author} {\bibfnamefont {S.}~\bibnamefont {Botti}},\ and\ \bibinfo {author} {\bibfnamefont {M.~A.~L.}\ \bibnamefont {Marques}},\ }\bibfield  {title} {\bibinfo {title} {Prediction of stable nitride perovskites},\ }\href {https://doi.org/10.1021/acs.chemmater.5b02026} {\bibfield  {journal} {\bibinfo  {journal} {Chem. Mater.}\ }\textbf {\bibinfo {volume} {27}},\ \bibinfo {pages} {5957} (\bibinfo {year} {2015})}\BibitemShut {NoStop}%
\bibitem [{\citenamefont {Gui}\ and\ \citenamefont {Dong}(2020)}]{PhysRevB.102.180103}%
  \BibitemOpen
  \bibfield  {author} {\bibinfo {author} {\bibfnamefont {C.}~\bibnamefont {Gui}}\ and\ \bibinfo {author} {\bibfnamefont {S.}~\bibnamefont {Dong}},\ }\bibfield  {title} {\bibinfo {title} {Pressure-induced ferroelectric phase of ${\mathrm{lamon}}_{3}$},\ }\href@noop {} {\bibfield  {journal} {\bibinfo  {journal} {Phys. Rev. B}\ }\textbf {\bibinfo {volume} {102}},\ \bibinfo {pages} {180103} (\bibinfo {year} {2020})}\BibitemShut {NoStop}%
\bibitem [{\citenamefont {Bell}(2008)}]{Bell2008}%
  \BibitemOpen
  \bibfield  {author} {\bibinfo {author} {\bibfnamefont {L.~E.}\ \bibnamefont {Bell}},\ }\bibfield  {title} {\bibinfo {title} {Cooling, heating, generating power, and recovering waste heat with thermoelectric systems},\ }\href@noop {} {\bibfield  {journal} {\bibinfo  {journal} {Science}\ }\textbf {\bibinfo {volume} {321}},\ \bibinfo {pages} {1457} (\bibinfo {year} {2008})}\BibitemShut {NoStop}%
\bibitem [{\citenamefont {Wu}\ and\ \citenamefont {Gao}(2018)}]{Wu2018}%
  \BibitemOpen
  \bibfield  {author} {\bibinfo {author} {\bibfnamefont {T.}~\bibnamefont {Wu}}\ and\ \bibinfo {author} {\bibfnamefont {P.}~\bibnamefont {Gao}},\ }\bibfield  {title} {\bibinfo {title} {Development of perovskite-type materials for thermoelectric application},\ }\href@noop {} {\bibfield  {journal} {\bibinfo  {journal} {Materials}\ }\textbf {\bibinfo {volume} {11}},\ \bibinfo {pages} {999} (\bibinfo {year} {2018})}\BibitemShut {NoStop}%
\bibitem [{\citenamefont {Zhao}\ \emph {et~al.}(2021)\citenamefont {Zhao}, \citenamefont {Zeng}, \citenamefont {Li}, \citenamefont {Lian}, \citenamefont {Dai}, \citenamefont {Meng},\ and\ \citenamefont {Ni}}]{Zhao2021}%
  \BibitemOpen
  \bibfield  {author} {\bibinfo {author} {\bibfnamefont {Y.}~\bibnamefont {Zhao}}, \bibinfo {author} {\bibfnamefont {S.}~\bibnamefont {Zeng}}, \bibinfo {author} {\bibfnamefont {G.}~\bibnamefont {Li}}, \bibinfo {author} {\bibfnamefont {C.}~\bibnamefont {Lian}}, \bibinfo {author} {\bibfnamefont {Z.}~\bibnamefont {Dai}}, \bibinfo {author} {\bibfnamefont {S.}~\bibnamefont {Meng}},\ and\ \bibinfo {author} {\bibfnamefont {J.}~\bibnamefont {Ni}},\ }\bibfield  {title} {\bibinfo {title} {Lattice thermal conductivity including phonon frequency shifts and scattering rates induced by quartic anharmonicity in cubic oxide and fluoride perovskites},\ }\href@noop {} {\bibfield  {journal} {\bibinfo  {journal} {Phys. Rev. B}\ }\textbf {\bibinfo {volume} {104}},\ \bibinfo {pages} {224304} (\bibinfo {year} {2021})}\BibitemShut {NoStop}%
\bibitem [{\citenamefont {Haque}\ \emph {et~al.}(2020)\citenamefont {Haque}, \citenamefont {Kee}, \citenamefont {Villalva}, \citenamefont {Ong},\ and\ \citenamefont {Baran}}]{Haque2020}%
  \BibitemOpen
  \bibfield  {author} {\bibinfo {author} {\bibfnamefont {M.~A.}\ \bibnamefont {Haque}}, \bibinfo {author} {\bibfnamefont {S.}~\bibnamefont {Kee}}, \bibinfo {author} {\bibfnamefont {D.~R.}\ \bibnamefont {Villalva}}, \bibinfo {author} {\bibfnamefont {W.-L.}\ \bibnamefont {Ong}},\ and\ \bibinfo {author} {\bibfnamefont {D.}~\bibnamefont {Baran}},\ }\bibfield  {title} {\bibinfo {title} {Halide perovskites: Thermal transport and prospects for thermoelectricity},\ }\href@noop {} {\bibfield  {journal} {\bibinfo  {journal} {Adv. Sci.}\ }\textbf {\bibinfo {volume} {7}},\ \bibinfo {pages} {1903389} (\bibinfo {year} {2020})}\BibitemShut {NoStop}%
\bibitem [{\citenamefont {Liu}\ \emph {et~al.}(2019)\citenamefont {Liu}, \citenamefont {Zhao}, \citenamefont {Li}, \citenamefont {Liu}, \citenamefont {Liscio}, \citenamefont {Milita}, \citenamefont {Schroeder},\ and\ \citenamefont {Fenwick}}]{Liu2019}%
  \BibitemOpen
  \bibfield  {author} {\bibinfo {author} {\bibfnamefont {T.}~\bibnamefont {Liu}}, \bibinfo {author} {\bibfnamefont {X.}~\bibnamefont {Zhao}}, \bibinfo {author} {\bibfnamefont {J.}~\bibnamefont {Li}}, \bibinfo {author} {\bibfnamefont {Z.}~\bibnamefont {Liu}}, \bibinfo {author} {\bibfnamefont {F.}~\bibnamefont {Liscio}}, \bibinfo {author} {\bibfnamefont {S.}~\bibnamefont {Milita}}, \bibinfo {author} {\bibfnamefont {B.~C.}\ \bibnamefont {Schroeder}},\ and\ \bibinfo {author} {\bibfnamefont {O.}~\bibnamefont {Fenwick}},\ }\bibfield  {title} {\bibinfo {title} {Enhanced control of self-doping in halide perovskites for improved thermoelectric performance},\ }\href@noop {} {\bibfield  {journal} {\bibinfo  {journal} {Nat. Commun.}\ }\textbf {\bibinfo {volume} {10}},\ \bibinfo {pages} {5750} (\bibinfo {year} {2019})}\BibitemShut {NoStop}%
\bibitem [{\citenamefont {van Roekeghem}\ \emph {et~al.}(2016)\citenamefont {van Roekeghem}, \citenamefont {Carrete}, \citenamefont {Oses}, \citenamefont {Curtarolo},\ and\ \citenamefont {Mingo}}]{Roekeghem2016}%
  \BibitemOpen
  \bibfield  {author} {\bibinfo {author} {\bibfnamefont {A.}~\bibnamefont {van Roekeghem}}, \bibinfo {author} {\bibfnamefont {J.}~\bibnamefont {Carrete}}, \bibinfo {author} {\bibfnamefont {C.}~\bibnamefont {Oses}}, \bibinfo {author} {\bibfnamefont {S.}~\bibnamefont {Curtarolo}},\ and\ \bibinfo {author} {\bibfnamefont {N.}~\bibnamefont {Mingo}},\ }\bibfield  {title} {\bibinfo {title} {High-throughput computation of thermal conductivity of high-temperature solid phases: The case of oxide and fluoride perovskites},\ }\href@noop {} {\bibfield  {journal} {\bibinfo  {journal} {Phys. Rev. X}\ }\textbf {\bibinfo {volume} {6}},\ \bibinfo {pages} {041061} (\bibinfo {year} {2016})}\BibitemShut {NoStop}%
\bibitem [{\citenamefont {Broido}\ \emph {et~al.}(2007)\citenamefont {Broido}, \citenamefont {Malorny}, \citenamefont {Birner}, \citenamefont {Mingo},\ and\ \citenamefont {Stewart}}]{10.1063/1.2822891}%
  \BibitemOpen
  \bibfield  {author} {\bibinfo {author} {\bibfnamefont {D.~A.}\ \bibnamefont {Broido}}, \bibinfo {author} {\bibfnamefont {M.}~\bibnamefont {Malorny}}, \bibinfo {author} {\bibfnamefont {G.}~\bibnamefont {Birner}}, \bibinfo {author} {\bibfnamefont {N.}~\bibnamefont {Mingo}},\ and\ \bibinfo {author} {\bibfnamefont {D.~A.}\ \bibnamefont {Stewart}},\ }\bibfield  {title} {\bibinfo {title} {Intrinsic lattice thermal conductivity of semiconductors from first principles},\ }\href@noop {} {\bibfield  {journal} {\bibinfo  {journal} {Appl. Phys. Lett.}\ }\textbf {\bibinfo {volume} {91}},\ \bibinfo {pages} {231922} (\bibinfo {year} {2007})}\BibitemShut {NoStop}%
\bibitem [{\citenamefont {Garg}\ \emph {et~al.}(2011)\citenamefont {Garg}, \citenamefont {Bonini}, \citenamefont {Kozinsky},\ and\ \citenamefont {Marzari}}]{PhysRevLett.106.045901}%
  \BibitemOpen
  \bibfield  {author} {\bibinfo {author} {\bibfnamefont {J.}~\bibnamefont {Garg}}, \bibinfo {author} {\bibfnamefont {N.}~\bibnamefont {Bonini}}, \bibinfo {author} {\bibfnamefont {B.}~\bibnamefont {Kozinsky}},\ and\ \bibinfo {author} {\bibfnamefont {N.}~\bibnamefont {Marzari}},\ }\bibfield  {title} {\bibinfo {title} {Role of disorder and anharmonicity in the thermal conductivity of {S}ilicon-{G}ermanium alloys: A first-principles study},\ }\href@noop {} {\bibfield  {journal} {\bibinfo  {journal} {Phys. Rev. Lett.}\ }\textbf {\bibinfo {volume} {106}},\ \bibinfo {pages} {045901} (\bibinfo {year} {2011})}\BibitemShut {NoStop}%
\bibitem [{\citenamefont {Feng}\ and\ \citenamefont {Ruan}(2016)}]{PhysRevB.93.045202}%
  \BibitemOpen
  \bibfield  {author} {\bibinfo {author} {\bibfnamefont {T.}~\bibnamefont {Feng}}\ and\ \bibinfo {author} {\bibfnamefont {X.}~\bibnamefont {Ruan}},\ }\bibfield  {title} {\bibinfo {title} {Quantum mechanical prediction of four-phonon scattering rates and reduced thermal conductivity of solids},\ }\href@noop {} {\bibfield  {journal} {\bibinfo  {journal} {Phys. Rev. B}\ }\textbf {\bibinfo {volume} {93}},\ \bibinfo {pages} {045202} (\bibinfo {year} {2016})}\BibitemShut {NoStop}%
\bibitem [{\citenamefont {Feng}\ \emph {et~al.}(2017)\citenamefont {Feng}, \citenamefont {Lindsay},\ and\ \citenamefont {Ruan}}]{PhysRevB.96.161201}%
  \BibitemOpen
  \bibfield  {author} {\bibinfo {author} {\bibfnamefont {T.}~\bibnamefont {Feng}}, \bibinfo {author} {\bibfnamefont {L.}~\bibnamefont {Lindsay}},\ and\ \bibinfo {author} {\bibfnamefont {X.}~\bibnamefont {Ruan}},\ }\bibfield  {title} {\bibinfo {title} {Four-phonon scattering significantly reduces intrinsic thermal conductivity of solids},\ }\href@noop {} {\bibfield  {journal} {\bibinfo  {journal} {Phys. Rev. B}\ }\textbf {\bibinfo {volume} {96}},\ \bibinfo {pages} {161201} (\bibinfo {year} {2017})}\BibitemShut {NoStop}%
\bibitem [{\citenamefont {Feng}\ and\ \citenamefont {Ruan}(2018)}]{PhysRevB.97.045202}%
  \BibitemOpen
  \bibfield  {author} {\bibinfo {author} {\bibfnamefont {T.}~\bibnamefont {Feng}}\ and\ \bibinfo {author} {\bibfnamefont {X.}~\bibnamefont {Ruan}},\ }\bibfield  {title} {\bibinfo {title} {Four-phonon scattering reduces intrinsic thermal conductivity of graphene and the contributions from flexural phonons},\ }\href@noop {} {\bibfield  {journal} {\bibinfo  {journal} {Phys. Rev. B}\ }\textbf {\bibinfo {volume} {97}},\ \bibinfo {pages} {045202} (\bibinfo {year} {2018})}\BibitemShut {NoStop}%
\bibitem [{\citenamefont {Kresse}\ and\ \citenamefont {Furthmüller}(1996{\natexlab{a}})}]{Kresse1996}%
  \BibitemOpen
  \bibfield  {author} {\bibinfo {author} {\bibfnamefont {G.}~\bibnamefont {Kresse}}\ and\ \bibinfo {author} {\bibfnamefont {J.}~\bibnamefont {Furthmüller}},\ }\bibfield  {title} {\bibinfo {title} {Efficient iterative schemes for ab initio total-energy calculations using a plane-wave basis set},\ }\href@noop {} {\bibfield  {journal} {\bibinfo  {journal} {Phys. Rev. B}\ }\textbf {\bibinfo {volume} {54}},\ \bibinfo {pages} {11169} (\bibinfo {year} {1996}{\natexlab{a}})}\BibitemShut {NoStop}%
\bibitem [{\citenamefont {Kresse}\ and\ \citenamefont {Furthmüller}(1996{\natexlab{b}})}]{Kresse1996b}%
  \BibitemOpen
  \bibfield  {author} {\bibinfo {author} {\bibfnamefont {G.}~\bibnamefont {Kresse}}\ and\ \bibinfo {author} {\bibfnamefont {J.}~\bibnamefont {Furthmüller}},\ }\bibfield  {title} {\bibinfo {title} {Efficiency of ab-initio total energy calculations for metals and semiconductors using a plane-wave basis set},\ }\href@noop {} {\bibfield  {journal} {\bibinfo  {journal} {Comput. Mater. Sci.}\ }\textbf {\bibinfo {volume} {6}},\ \bibinfo {pages} {15} (\bibinfo {year} {1996}{\natexlab{b}})}\BibitemShut {NoStop}%
\bibitem [{\citenamefont {Bl\"ochl}(1994)}]{Blochl1994}%
  \BibitemOpen
  \bibfield  {author} {\bibinfo {author} {\bibfnamefont {P.~E.}\ \bibnamefont {Bl\"ochl}},\ }\bibfield  {title} {\bibinfo {title} {Projector augmented-wave method},\ }\href@noop {} {\bibfield  {journal} {\bibinfo  {journal} {Phys. Rev. B}\ }\textbf {\bibinfo {volume} {50}},\ \bibinfo {pages} {17953} (\bibinfo {year} {1994})}\BibitemShut {NoStop}%
\bibitem [{\citenamefont {Perdew}\ \emph {et~al.}(1996)\citenamefont {Perdew}, \citenamefont {Burke},\ and\ \citenamefont {Ernzerhof}}]{perdew1996generalized}%
  \BibitemOpen
  \bibfield  {author} {\bibinfo {author} {\bibfnamefont {J.~P.}\ \bibnamefont {Perdew}}, \bibinfo {author} {\bibfnamefont {K.}~\bibnamefont {Burke}},\ and\ \bibinfo {author} {\bibfnamefont {M.}~\bibnamefont {Ernzerhof}},\ }\bibfield  {title} {\bibinfo {title} {Generalized gradient approximation made simple},\ }\href@noop {} {\bibfield  {journal} {\bibinfo  {journal} {Phys. Rev. Lett.}\ }\textbf {\bibinfo {volume} {77}},\ \bibinfo {pages} {3865} (\bibinfo {year} {1996})}\BibitemShut {NoStop}%
\bibitem [{\citenamefont {Togo}\ and\ \citenamefont {Tanaka}(2015)}]{Togo2015}%
  \BibitemOpen
  \bibfield  {author} {\bibinfo {author} {\bibfnamefont {A.}~\bibnamefont {Togo}}\ and\ \bibinfo {author} {\bibfnamefont {I.}~\bibnamefont {Tanaka}},\ }\bibfield  {title} {\bibinfo {title} {First principles phonon calculations in materials science},\ }\href {https://doi.org/10.1016/j.scriptamat.2015.07.021} {\bibfield  {journal} {\bibinfo  {journal} {Scr. Mater.}\ }\textbf {\bibinfo {volume} {108}},\ \bibinfo {pages} {1} (\bibinfo {year} {2015})}\BibitemShut {NoStop}%
\bibitem [{\citenamefont {Esfarjani}\ and\ \citenamefont {Stokes}(2008)}]{Esfarjani2008}%
  \BibitemOpen
  \bibfield  {author} {\bibinfo {author} {\bibfnamefont {K.}~\bibnamefont {Esfarjani}}\ and\ \bibinfo {author} {\bibfnamefont {H.~T.}\ \bibnamefont {Stokes}},\ }\bibfield  {title} {\bibinfo {title} {Method to extract anharmonic force constants from first principles calculations},\ }\href {https://doi.org/10.1103/PhysRevB.77.144112} {\bibfield  {journal} {\bibinfo  {journal} {Phys. Rev. B}\ }\textbf {\bibinfo {volume} {77}},\ \bibinfo {pages} {144112} (\bibinfo {year} {2008})}\BibitemShut {NoStop}%
\bibitem [{\citenamefont {Wang}\ \emph {et~al.}(2010)\citenamefont {Wang}, \citenamefont {Wang}, \citenamefont {Wang}, \citenamefont {Mei}, \citenamefont {Shang}, \citenamefont {Chen},\ and\ \citenamefont {Liu}}]{Wang2010}%
  \BibitemOpen
  \bibfield  {author} {\bibinfo {author} {\bibfnamefont {Y.}~\bibnamefont {Wang}}, \bibinfo {author} {\bibfnamefont {J.~J.}\ \bibnamefont {Wang}}, \bibinfo {author} {\bibfnamefont {W.~Y.}\ \bibnamefont {Wang}}, \bibinfo {author} {\bibfnamefont {Z.~G.}\ \bibnamefont {Mei}}, \bibinfo {author} {\bibfnamefont {S.~L.}\ \bibnamefont {Shang}}, \bibinfo {author} {\bibfnamefont {L.~Q.}\ \bibnamefont {Chen}},\ and\ \bibinfo {author} {\bibfnamefont {Z.~K.}\ \bibnamefont {Liu}},\ }\bibfield  {title} {\bibinfo {title} {A mixed-space approach to first-principles calculations of phonon frequencies for polar materials},\ }\href {https://doi.org/10.1088/0953-8984/22/20/202201} {\bibfield  {journal} {\bibinfo  {journal} {J. Phys.: Condens. Matter}\ }\textbf {\bibinfo {volume} {22}},\ \bibinfo {pages} {202201} (\bibinfo {year} {2010})}\BibitemShut {NoStop}%
\bibitem [{\citenamefont {Li}\ \emph {et~al.}(2014)\citenamefont {Li}, \citenamefont {Carrete}, \citenamefont {Katcho},\ and\ \citenamefont {Mingo}}]{Li2014}%
  \BibitemOpen
  \bibfield  {author} {\bibinfo {author} {\bibfnamefont {W.}~\bibnamefont {Li}}, \bibinfo {author} {\bibfnamefont {J.}~\bibnamefont {Carrete}}, \bibinfo {author} {\bibfnamefont {N.~A.}\ \bibnamefont {Katcho}},\ and\ \bibinfo {author} {\bibfnamefont {N.}~\bibnamefont {Mingo}},\ }\bibfield  {title} {\bibinfo {title} {{ShengBTE}: A solver of the boltzmann transport equation for phonons},\ }\href {https://doi.org/10.1016/j.cpc.2014.02.015} {\bibfield  {journal} {\bibinfo  {journal} {Comput. Phys. Commun.}\ }\textbf {\bibinfo {volume} {185}},\ \bibinfo {pages} {1747} (\bibinfo {year} {2014})}\BibitemShut {NoStop}%
\bibitem [{\citenamefont {Han}\ \emph {et~al.}(2022)\citenamefont {Han}, \citenamefont {Yang}, \citenamefont {Li}, \citenamefont {Feng},\ and\ \citenamefont {Ruan}}]{Han2022}%
  \BibitemOpen
  \bibfield  {author} {\bibinfo {author} {\bibfnamefont {Z.}~\bibnamefont {Han}}, \bibinfo {author} {\bibfnamefont {X.}~\bibnamefont {Yang}}, \bibinfo {author} {\bibfnamefont {W.}~\bibnamefont {Li}}, \bibinfo {author} {\bibfnamefont {T.}~\bibnamefont {Feng}},\ and\ \bibinfo {author} {\bibfnamefont {X.}~\bibnamefont {Ruan}},\ }\bibfield  {title} {\bibinfo {title} {{FourPhonon}: An extension module to shengbte for computing four-phonon scattering rates and thermal conductivity},\ }\href {https://doi.org/10.1016/j.cpc.2021.108179} {\bibfield  {journal} {\bibinfo  {journal} {Comput. Phys. Commun.}\ }\textbf {\bibinfo {volume} {270}},\ \bibinfo {pages} {108179} (\bibinfo {year} {2022})}\BibitemShut {NoStop}%
\bibitem [{\citenamefont {Singh}\ and\ \citenamefont {Tripathi}(2018)}]{10.1063/1.5035135}%
  \BibitemOpen
  \bibfield  {author} {\bibinfo {author} {\bibfnamefont {S.}~\bibnamefont {Singh}}\ and\ \bibinfo {author} {\bibfnamefont {M.~N.}\ \bibnamefont {Tripathi}},\ }\bibfield  {title} {\bibinfo {title} {{Sr}-doped {LaMoN$_3$} and {LaWN$_3$}: New degenerate p-type nitrides},\ }\href@noop {} {\bibfield  {journal} {\bibinfo  {journal} {J. Appl. Phys.}\ }\textbf {\bibinfo {volume} {124}},\ \bibinfo {pages} {065109} (\bibinfo {year} {2018})}\BibitemShut {NoStop}%
\bibitem [{\citenamefont {Grosso}\ \emph {et~al.}(2023)\citenamefont {Grosso}, \citenamefont {Davies}, \citenamefont {Zhu}, \citenamefont {Walsh},\ and\ \citenamefont {Scanlon}}]{D3SC02171H}%
  \BibitemOpen
  \bibfield  {author} {\bibinfo {author} {\bibfnamefont {B.~F.}\ \bibnamefont {Grosso}}, \bibinfo {author} {\bibfnamefont {D.~W.}\ \bibnamefont {Davies}}, \bibinfo {author} {\bibfnamefont {B.}~\bibnamefont {Zhu}}, \bibinfo {author} {\bibfnamefont {A.}~\bibnamefont {Walsh}},\ and\ \bibinfo {author} {\bibfnamefont {D.~O.}\ \bibnamefont {Scanlon}},\ }\bibfield  {title} {\bibinfo {title} {Accessible chemical space for metal nitride perovskites},\ }\href@noop {} {\bibfield  {journal} {\bibinfo  {journal} {Chem. Sci.}\ }\textbf {\bibinfo {volume} {14}},\ \bibinfo {pages} {9175} (\bibinfo {year} {2023})}\BibitemShut {NoStop}%
\bibitem [{\citenamefont {Slack}(1979)}]{slack1979thermal}%
  \BibitemOpen
  \bibfield  {author} {\bibinfo {author} {\bibfnamefont {G.~A.}\ \bibnamefont {Slack}},\ }\bibfield  {title} {\bibinfo {title} {Thermal conductivity of pure and impure materials},\ }\href@noop {} {\bibfield  {journal} {\bibinfo  {journal} {Solid State Phys.}\ }\textbf {\bibinfo {volume} {34}},\ \bibinfo {pages} {1} (\bibinfo {year} {1979})}\BibitemShut {NoStop}%
\bibitem [{\citenamefont {Ziman}(1960)}]{ziman1960electrons}%
  \BibitemOpen
  \bibfield  {author} {\bibinfo {author} {\bibfnamefont {J.~M.}\ \bibnamefont {Ziman}},\ }\href@noop {} {\emph {\bibinfo {title} {Electrons and Phonons: The Theory of Transport Phenomena in Solids}}}\ (\bibinfo  {publisher} {Oxford University Press},\ \bibinfo {year} {1960})\BibitemShut {NoStop}%
\bibitem [{\citenamefont {Zeng}\ \emph {et~al.}(2021)\citenamefont {Zeng}, \citenamefont {Zhang}, \citenamefont {Yu}, \citenamefont {Li}, \citenamefont {Pei},\ and\ \citenamefont {Chen}}]{Zeng2023}%
  \BibitemOpen
  \bibfield  {author} {\bibinfo {author} {\bibfnamefont {Z.}~\bibnamefont {Zeng}}, \bibinfo {author} {\bibfnamefont {C.}~\bibnamefont {Zhang}}, \bibinfo {author} {\bibfnamefont {H.}~\bibnamefont {Yu}}, \bibinfo {author} {\bibfnamefont {W.}~\bibnamefont {Li}}, \bibinfo {author} {\bibfnamefont {Y.}~\bibnamefont {Pei}},\ and\ \bibinfo {author} {\bibfnamefont {Y.}~\bibnamefont {Chen}},\ }\bibfield  {title} {\bibinfo {title} {Ultralow and glass-like lattice thermal conductivity in crystalline {BaAg$_{2}$Te$_{2}$}: Strong fourth-order anharmonicity and crucial diffusive thermal transport},\ }\href {https://doi.org/10.1016/j.mtphys.2021.100487} {\bibfield  {journal} {\bibinfo  {journal} {Mater. Today Phys.}\ }\textbf {\bibinfo {volume} {21}},\ \bibinfo {pages} {100487} (\bibinfo {year} {2021})}\BibitemShut {NoStop}%
\bibitem [{\citenamefont {Yang}\ \emph {et~al.}(2019)\citenamefont {Yang}, \citenamefont {Feng}, \citenamefont {Li},\ and\ \citenamefont {Ruan}}]{Yang2019}%
  \BibitemOpen
  \bibfield  {author} {\bibinfo {author} {\bibfnamefont {X.}~\bibnamefont {Yang}}, \bibinfo {author} {\bibfnamefont {T.}~\bibnamefont {Feng}}, \bibinfo {author} {\bibfnamefont {J.}~\bibnamefont {Li}},\ and\ \bibinfo {author} {\bibfnamefont {X.}~\bibnamefont {Ruan}},\ }\bibfield  {title} {\bibinfo {title} {Stronger role of four-phonon scattering than three-phonon scattering in thermal conductivity of iii-v semiconductors at room temperature},\ }\href {https://doi.org/10.1103/PhysRevB.100.245203} {\bibfield  {journal} {\bibinfo  {journal} {Phys. Rev. B}\ }\textbf {\bibinfo {volume} {100}},\ \bibinfo {pages} {245203} (\bibinfo {year} {2019})}\BibitemShut {NoStop}%
\end{thebibliography}%

\end{document}